\documentclass[twocolumn,showpacs]{revtex4}

\usepackage{graphicx}
\usepackage{amsmath}
\usepackage{epsfig}
\usepackage{amssymb}
\usepackage{theorem}

\usepackage{hhline}

\theoremstyle{plain}

\theoremstyle{plain}

\numberwithin{equation}{section}

\newcommand{\ket}[1]{\left\vert#1\right\rangle}

\newcommand{\beq}{\begin{equation}}
\newcommand{\eeq}{\end{equation}}
\newcommand{\bea}{\begin{eqnarray}}
\newcommand{\eea}{\end{eqnarray}}

\newcommand{\tr}{\mbox{Tr}}

\makeatletter
\def\btt#1{\texttt{\@backslashchar#1}}
\DeclareRobustCommand\bblash{\btt{\@backslashchar}}
\makeatother

\begin{document}

\title{Casimir Invariants for Systems Undergoing Collective Motion}

\author{C. Allen Bishop$^{1}$}\author{Mark S. Byrd$^{1,2}$}\author{Lian-Ao Wu$^{3,4}$}

\affiliation{$^1$Physics Department, Southern Illinois University, 
Carbondale, Illinois 62901-4401}
\affiliation{$^2$Computer Science Department, Southern Illinois University, 
Carbondale, Illinois 62901}
\affiliation{$^3$Department of Theoretical Physics and History of
  Science, The Basque Country University (EHU/UPV), P.O. Box 644, 48080
  Bilbao, Spain}
\affiliation{$^4$IKERBASQUE--Basque Foundation for Science, 48011,
  Bilbao, Spain}
\date{\today}

\begin{abstract}
Dicke states are states of a collection of particles which have been
under active investigation for several reasons.  One reason is that
the decay rates of these states 
can be quite different from a set of independently
evolving particles.  Another reason is that a particular class of
these states are decoherence-free or noiseless with respect to a set 
of errors.  These noiseless states, or more generally subsystems, 
can avoid certain types of errors in quantum information processing
devices.  Here we provide a method for calculating invariants of
systems of particles undergoing collective motions.  These invariants 
can be used to determine a complete set of commuting observables for
a class of Dicke states as well as 
identify possible logical operations for decoherence-free/noiseless
subsystems.  Our method is quite general and provides results for cases
where the constituent particles have more than two internal states.  
\end{abstract}


\maketitle




\section{Introduction}
Decoherence-free/noiseless subsystems (DFS) are now part of an
arsenal of weapons used to prevent errors in quantum information
processing and storage 
\cite{Zanardi:97c,Duan:98,Lidar:PRL98,Knill:99a,Kempe:00,Lidar:00a}.     
(For reviews see \cite{Lidar/Whaley:03,Byrd/etal:pqe04}.)  DFS are
subsystems which are immune to certain types of errors.  The most
common type found in the literature is a DFS which is immune to
collective errors.  These types of quantum systems were studied 
earlier by Dicke in a different context \cite{Dicke:54}.  

There are several types of states which are now called Dicke states. 
One such set corresponds to a set of particles which undergo a
collective motion, are distinguishable, and do not interact with each
other.  These states are unchanged by particle interchange, or more
generally, the interchange of particular constituents \cite{partnote}.
One particularly clear example is a gas interacting with an external
field which has a wavelength significantly longer than the container
confining the particles.  These are also conditions for collective
motion, i.e., the external field 
interacts in the same way with each particle.  In this case, if the
size of the container $\sim R$ and the wavelength of the field is
$\lambda$, then the ``Dicke limit'' $\lambda \gg R$ is said to be
satisfied.  In this limit, when the external 
influence gives rise to errors in a quantum computing device, the
errors are called collective, whether they describe an evolution of
each particle which is unitary or not.  

Since errors are the greatest obstacle to building a fully functional
quantum computing device, any method which aids in the prevention of
errors is quite important.  However, for the practical use of a DFS/NS
for quantum information processing one requires the ability to perform
universal computing on these states.  This requires finding evolutions
which do not take the states out of the protected subspace during
gating operations \cite{Kempe:00}.  We refer to such operations as
being compatible with the DFS structure.  In the physical systems
considered by Dicke, one could imagine evolutions of the states which
do not change the essential features of the state (energy or total
angular momentum quantum numbers), but are indeed nontrivial
evolutions.  In the case of quantum information processing, these
enable quantum computing in a DFS.  

In both the early analysis of Dicke states and also quantum computing
applications, primarily only two internal states of the constituents
were considered.  However, three or more internal states of an atom
could certainly become important in various experiments and could also
arise in particle physics where more than two degrees of freedom are
associated with both flavor and color symmetries.  Recent
experiments
\cite{Prevedel/etal:09,Wieczorek/etal:09,Harkonen/etal:09} 
and proposed experiments 
\cite{Kiffner/etal:07,Bishop/Byrd:08,Hume/etal:09}
have provided explicit constructions for these so-called
Dicke states using a variety of physical systems.  

Here we carry the consideration of Dicke states to the extreme.  We 
consider collections of particles undergoing some collective motions, 
for example collective errors, and ask the following question.  What 
Hamiltonians give rise to evolutions which are compatible with these 
motions?  Our results are not restricted to any particular number of
internal states  for each of the constituents, nor are they restricted
to any number of particles.  We then answer the question by using a
construction of invariants analogous to Casimir's construction of
invariants for Lie algebras and Lie groups.  

In Section \ref{sec:casops} we review the standard Casimir
construction for single-particle invariants.  In Section 
\ref{sec:manyptclecas} we extend
the construction to sets of $N$ particles each with
d internal states.  Section \ref{sec:phys} discusses the physical
implications of our results.  In particular, we discuss the
use of these invariants for Dicke state identification as well as 
the manipulation of decoherence-free or noiseless subsystem.
Section \ref{sec:concl} concludes.


\section{Identifying Invariants}

\label{sec:casops}

A Casimir Operator is a member of the center of the universal
enveloping algebra meaning such an operator will 
commute with every element of the universal enveloping
algebra.  For matrix representations of quantum evolutions, which we
will consider here, the universal enveloping algebra is the algebra of
all products of Lie algebra basis elements.  It is most important for
our purposes that the Casimir operators commute with every generator
of the Lie algebra and the collective
errors form a representation of the Lie algebra (which is the algebra
of Hermitian matrices).  Once we find such invariants, we will have
the set of Hamiltonians which commute with collective errors and are
therefore compatible transformations.  We begin by reviewing the
construction of Casimir invariants.  

Let a basis for the Lie algebra of SU(d) (hereafter denoted 
${\cal L}$(SU(d))) be given by a set $\{\lambda_i \}$ with the
normalization and properties described in the Appendix.  The Casimir
operators of SU(d) are known.  The most familiar, the quadratic
Casimir, is proportional to the sum of the squares of the
elements, 
\begin{equation}
\label{eq:quadcasI}
C_2 \propto \sum_i \lambda_i \lambda_i.
\end{equation}
This along with all other Casimir operators can be obtained using the
formula \cite{Gruber:64,Fuchs/Schweigert} 
\begin{equation}
\label{eq:In}
I_n = \tr(\mathrm{ad}_{\lambda_{a_1}} \circ \mathrm{ad}_{\lambda_{a_2}} \circ
\cdots \circ\mathrm{ad}_{\lambda_{a_n}}) \lambda_{a_1}\lambda_{a_2} ... 
\lambda_{a_n}.
\end{equation}

For example, 
\begin{equation}
C_2 = \sum_{a_1,a_2,b_1,b_2} f_{a_1,b_1,b_2}f_{a_2,b_2,b_1}\lambda_{a_1}\lambda_{a_2},
\end{equation}
which reduces to Eq.~(\ref{eq:quadcasI}) using Eq.~(\ref{eq:Mac(2.12)}).
It turns out that the formula given in Eq.~(\ref{eq:In}) does not
produce independent invariants for the collective errors.  However,
the independent invariants can be obtained \cite{Gruber:64} and may be
written in terms of the totally symmetric $d$-tensor.  For example, 
the cubic Casimir invariant is 
\begin{equation}
C_3 = \sum_{ijk} d_{ijk}\lambda_i \lambda_j \lambda_k.
\end{equation}
Higher order Casimir operators can be constructed using the general
formulation
\begin{eqnarray}
C_n &=& \!\!\!\!\sum_{i_1,i_2,...,i_n} d_{i_1,i_2,i_3}d_{i_3,i_4,i_5}, \dots
d_{i_{n-4},i_{n-3},i_{n-2}} \nonumber \\
&& \;\;\;\;\; \times d_{i_{n-2},i_{n-1},i_{n}}\lambda_{i_1}\lambda_{i_2}\lambda_{i_4}\dots\lambda_{i_{n-1}}\lambda_{i_n}.
\end{eqnarray}
The sum is over all elements of the algebra.  

To show independence, one may begin with Eq.~(\ref{eq:In}) and
reduce the expressions using the identities in the appendix.  Here our
objective is to find a set of operators which commute with the set of
collective motions.  A basis for these motions is given by the set of
operators of the form
\begin{equation}
\label{eq:S}
S_j = \sum_\alpha \lambda^{(\alpha)}_{j},
\end{equation}
where the sum is taken over the particles in the system.  These types of
operators also form a basis for the collective errors acting on a DFS/NS and
linear combinations give the stabilizer elements.  (See Sec.~%
\ref{sec:DFSstabilizer} for the definition and discussion.)  
An element of the algebra (with real coefficients) which commutes with
these provides the Hamiltonians which are compatible with a DFS/NS.


\section{Explicit forms for the Invariants} 

\label{sec:manyptclecas}

In this section we will find a set of independent operations  
for which each element of the set commutes with 
all members of the algebra formed by the $S_j$.  Denote the 
algebra of the $S_j$ by ${\cal A}$.  

Note that the Casimir operators 
formed from the elements $S_j$ form 
a representation of ${\cal L}$(SU(d)) if the $\lambda_i$ do 
\cite{Byrd:06}.  Therefore these are 
invariants of the algebra ${\cal A}$, i.e. they commute with elements
of this algebra.  However, this is not an
irreducible algebra. Thus the construction must rely on the
identification of the irreducible components.  

To proceed, we first calculate the Casimir invariants of 
${\cal L}$(SU(d)).  Then, noting that linear combinations of these
invariants are also invariants, we extract reducible components of the
invariants.  From a physical perspective, this means identifying
n-body interactions which are contained within the m-body interactions
where n$\leq$m.  

The quadratic Casimir operator for the algebra ${\cal A}$ is 
\begin{equation}
J_2 = \sum_{i,j,k,l} f_{ijk}f_{kli}S_j S_l \propto \sum_j S_jS_j.  
\end{equation}
Expanding this in terms of the basis elements $\{\lambda_i\}$
gives
\begin{eqnarray}
\label{eq:J2}
J_2 &\propto& \sum_i \left(\sum_{\alpha}\lambda_i^{(\alpha)}\right)^2\nonumber\\
    &=&  \sum_i \left( \sum_\alpha (\lambda_i^{(\alpha)})^2 
        + 2 \sum_{\alpha<\beta}\lambda_i^{(\alpha)}\lambda_i^{(\beta)} \right). 
\end{eqnarray}
Note that the first term of the last expression
is the sum of single-particle Casimir invariants. 
This allows us to infer 
that the second term in Eq.~(\ref{eq:J2}) is also an invariant quantity. 
Furthermore, the only nontrivial contributions appearing in the 
commutator $\left[\sum_i \lambda_i^{(\alpha)}\lambda_i^{(\beta)} , S_l \right]$ 
have the form
\begin{equation}
\label{eq:comm}
[\lambda_i^{(\alpha)},\lambda_j^{(\alpha)}]\lambda_i^{(\beta)} + 
\lambda_i^{(\alpha)}[\lambda_i^{(\beta)},\lambda_j^{(\beta)}],
\end{equation}
with all other terms vanishing. Since this can be rewritten as  
\begin{equation}
2if_{ijk}(\lambda_k^{(\alpha)} \lambda_i^{(\beta)} - 
\lambda_k^{(\alpha)} \lambda_i^{(\beta)}) = 0,
\end{equation}
we find that
\begin{equation}
\label{eq:genex}
I_2^{(\alpha,\beta)} = \sum_i \lambda_i^{(\alpha)}\lambda_i^{(\beta)}
\end{equation}
is also an independent invariant for each pair $(\alpha,\beta)$.

Now consider 
\begin{eqnarray}
J_3\!\!\! &=&\!\!\! \sum f_{ijk}f_{klm}f_{mni} S_j S_l S_n \nonumber \\
    &=& \!\!\!  \sum  f_{ijk}f_{klm}f_{mni}\! \left(\sum_\alpha
      \lambda_j^{(\alpha)}\right)\!\!
             \left(\sum_\beta \lambda_l^{(\beta)}\right)\!\!
             \left(\sum_\gamma \lambda_n^{(\gamma)}\right). \nonumber \\
\end{eqnarray}
Expanding the sums over the particle (Greek) indices, 
and reducing the
results, three types of terms are obtained.  First, if all three
superscripts are the same, for example
$\lambda_i^{(\alpha)}\lambda_j^{(\alpha)}\lambda_k^{(\alpha)}$,
the term reduces to the quadratic Casimir invariant for particle
$\alpha$.  Since any linear combination of invariants is invariant,
the sum of all terms having this form is also invariant.
Second, if two are the same, 
e.g. $\lambda_i^{(\alpha)}\lambda_j^{(\alpha)}\lambda_k^{(\beta)}$, 
then the result reduces to $I_2^{(\alpha,\beta)}$, thus terms of this
form are also invariant quantities.  Third, if all three are
different, we obtain 
\begin{equation}
I_3^{(\alpha,\beta,\gamma)} =
\sum_{ijk}f_{ijk}\lambda_i^{(\alpha)}\lambda_j^{(\beta)}\lambda_k^{(\gamma)}, 
\end{equation}
as an independent invariant.  Notice this case is different from the
ordinary Casimir construction where no such independent invariant
arises for a term of the form of $J_3$. 

Defining and 
expanding $J_4$ produces one new invariant,
\begin{equation}
I_4^{(\alpha,\beta,\gamma)} =
\sum_{ijk}d_{ijk}\lambda_i^{(\alpha)}\lambda_j^{(\beta)}\lambda_k^{(\gamma)}.  
\end{equation}
Continuing with this will iteratively produce a set of independent
invariants for collective motions of particles.  
For three qutrits this set, $I_2, I_3, I_4$ is complete 
\cite{Bishop/Byrd:09}.


\section{Physical Implications} 

\label{sec:phys}

After the motivation in the introduction and the construction of the
invariants, we now consider more explicitly the implications of our
findings.


\subsection{Motion of Dicke States}

In Ref.~\cite{Dicke:54} Dicke examined the spontaneous radiation 
of photons emitted from a gas consisting of two-level particles. 
Gasses of both small and large extent were treated separately, the 
scale being determined relative to the wavelength $\lambda$ of an 
externally applied field.  Taking $R$ to be the spatial extent of the
container, the two cases correspond to $\lambda \gg
R$ or $\lambda \ll R$.  In both cases it was assumed that there 
was insufficient overlap of the wave functions of separate particles to 
require symmetrization of the states.  It was also assumed that 
each particle coupled to the common radiation field via an electric dipole 
interaction. In general, the interaction energy of the $\alpha$th particle 
with the field can be written as 
\begin{equation}
H_I^{(\alpha)} = -{\bf{A(r_{\it{\alpha}})}} \cdot  
({\bf{e_1}}\sigma_x^{(\alpha)} + {\bf{e_2}}\sigma_y^{(\alpha)}),
\end{equation}
for some constant real vectors ${\bf{e_1}}$ and ${\bf{e_2}}$. 

In the case of a gas confined to a small region of space the  
vector potential can effectively be considered an  
independent function of the spatial coordinates ${\bf{r_{\it{\alpha}}}}$. 
In this approximation the total interaction energy becomes 
\begin{equation}
H_I = c_1 \sum_\alpha \sigma_x^{(\alpha)} +  c_2 \sum_\alpha \sigma_y^{(\alpha)},
\end{equation}
where $c_1$ and $c_2$ denote constants. There are two degrees 
of freedom associated with the internal energy 
of any given particle. The energy eigenvalues of the 
$j$th particle, corresponding to the diagonal operator 
$\sigma_z^{(\alpha)}$, take on the values $\pm  \hbar \omega /2$. The sum 
of all internal particle energies, together with the 
translational energy of the gas $H_0$ and the interaction with the 
field, provides a complete description of a gaseous system 
consisting of mutually noninteracting particles. 

The Hamiltonian for this system can be broken up into two parts,
\begin{equation}
\label{eq:DickeHam}
H = H_0 + \left(c_1 \sum_\alpha \sigma_x^{(\alpha)} +  c_2 \sum_\alpha \sigma_y^{(\alpha)} 
+  \hbar \omega /2  \sum_\alpha \sigma_z^{(\alpha)} \right),
\end{equation}
where the first part describes the translational energy of the system
and thus depends solely on the spatial positions 
${\bf{r_{\it{\alpha}}}}$ while the second is a quantity independent of 
these coordinates. As a result, these two parts commute implying 
the existence of simultaneous eigenfunctions of the two 
contributions. Let us denote these energy eigenstates 
\begin{equation}
\psi_{pq} = U_p({\bf{r}}_1,{\bf{r}}_2, \hdots , {\bf{r}}_N)\:\Phi_{q},
\end{equation}  
where $U_p$ depends on the spatial coordinates and 
$\Phi_{q}$ is a function of the internal coordinates. The 
operators $S_i = \sum_\alpha\sigma_i^{(\alpha)} \; (i=x,y,z)$ not only
individually 
commute with the spatially independent quantity $S^2 = S_x^2 + S_y^2 + S_z^2$, 
but also satisfy the same commutation relations (up to 
a multiplicative scaling factor) as the three components 
of angular momentum.  In other words, they form a representation of the
SO(3) algebra.  Stationary states of this system can therefore 
be identified with those eigenstates that conserve the square of the 
total angular momentum operator, i.e., $\Phi_{q} \equiv \Phi_{jm}$, 
with $S^2 \Phi_{jm} = j(j+1)\Phi_{jm}$ and $|m| \leq j \leq N/2$. 
Consequently, the stationary states of a gaseous system confined to a small 
region of space can be expressed as
\begin{equation}
\psi_{pjm} = U_p({\bf{r}}_1,{\bf{r}}_2, \hdots , {\bf{r}}_N)\:\Phi_{jm}.
\end{equation}

Since the individual particles which form the gas all experience 
a common interaction with the radiation field, the system as a whole 
evolves in a collective manner.  However, while this collective 
motion is occurring on these states, they may still undergo other
non-trivial evolutions.  Such operations conserve energy and angular
momentum.  Hamiltonians corresponding to these non-trivial
evolutions commute with the collective operators and thus can be
constructed from the previously derived invariants.  Furthermore, the
number of internal states is not restricted to two, but can be
arbitrary.  Many internal states may be undergoing simultaneous
transitions to other internal states, collectively, while still
undergoing this evolution.

In the next section we consider a particular type of Dicke state
which is actually invariant under these collective motions.  
Although the argument follows 
the usual treatment regarding the compatibility of transformations 
of a collective DFS/NS, it applies to the present case as well 
since DFS/NS states suitable for quantum information processing 
correspond to degenerate Dicke states.


\subsection{DFS-Compatible Hamiltonians} 

\label{sec:DFSstabilizer}

Let us suppose that the Dicke states corresponding to a 
collective DFS/NS are spanned by the set $\{ \ket{\lambda} \otimes 
\ket{\mu}\}$, with $\lambda = 1, \hdots , d$ and 
$\mu = 1, \hdots , n$. Here the $\ket{\lambda}$'s distinguish 
a particular basis state of an encoded $d$-state system  
and the $\ket{\mu}$'s label the $n$ orthogonal 
elements which span each qudit dimension. When acted 
upon by the collective errors $S_{j}$ these DFS/NS 
states have the property that 
\begin{equation}
S_{j} \ket{\lambda} \otimes \ket{\mu} = 
\sum_{\mu^{\prime}=1}^n M_{\mu \mu^{\prime}, j} \ket{\lambda} \otimes 
\ket{\mu^{\prime}}.
\end{equation}
In other words, these encoded qudit states remain unaffected by the 
presence of such noise since they map every $\ket{\lambda}$ 
to itself. One can parameterize the collective errors using 
a set of time-independent complex numbers $\{ v_{j} \}$,
\begin{equation}
D(v_1,v_2,...) = \exp \left[ \sum_{j} v_{j} S_{j} \right].
\end{equation}

The DFS/NS states are not the only accessible states inherent to a 
system. There are some orthogonal to these which cannot protect 
against collective noise. When information is leaked into these 
regions of the systems Hilbert space it may be permanently lost. 
Gates which are used to 
manipulate the state of an encoded qudit 
should therefore operate in a manner such that they map DFS/NS states 
to other DFS/NS states. It can be shown that a sufficient condition 
for a transformation 
$U = \exp(-iHt)$ to satisfy this compatibility requirement is that 
\begin{equation}
UD(v_1,v_2,\hdots)U^{\dagger} = D(v_1^{\prime},v_2^{\prime},\hdots),
\end{equation}
or, equivalently
\begin{equation}
\sum_{j} v_{j} U S_{j} U^{\dagger} = 
\sum_{j} v_{j}^{\prime} S_{j}.
\end{equation}
Taking the derivative of both sides of this equation with respect 
to time yields a sufficient condition for a Hamiltonian 
to generate a compatible transformation
\begin{equation}
[H,S_{j}] = 0, \:\:\:\: \forall S_{j}. 
\end{equation}
Since the Casimir operators for the algebra $\cal{A}$ satisfy 
this condition, they can be used to generate nondissapative 
transformations of a DFS/NS encoding. We will discuss the implications 
of these results for the case 
of a three qudit encoding next, with a particular emphasis on 
the ability of these operations to generate universal quantum computation.


\subsection{Three Qudits} 

\label{sec:cdits}

As mentioned earlier, a basis for the collective
errors is given by the set
\begin{equation}
S_i = \sum_\alpha\lambda_i^{(\alpha)},
\end{equation}
where the subscript indicates the type of error and the superscript
labels the particle on which the operator acts. The invariants 
$I_2$, $I_3$, and $I_4$ not only commute with every element of this 
set, but can also be used to form a representation 
of the Lie algebra of $SU(2)$ \cite{Bishop/Byrd:09}. It has been shown 
that the encoded, or logical analogues of the Pauli matrices acting on an 
encoded qubit can be given in terms of these invariants by the relations 
\begin{equation}
\bar{X} = \frac{1}{2\sqrt{3}}\left[I_2^{(2,3)}-I_2^{(1,3)}\right], \:\: 
\bar{Y} = \frac{I_3}{2\sqrt{3}},
\end{equation}
and
\begin{equation}
\bar{Z} = \left[I_2^{(2,3)}+ I_2^{(1,3)} -2I_2^{(1,2)}  \right]/6.
\end{equation}
All three of these generators can be expressed in terms 
of two body interactions since $I_3$ can be 
decomposed into products of $I_2$. In fact, the invariant $I_2$ alone suffices 
to perform universal computation using encoded qubits that are comprised 
of three physical qudits since they are able to generate any single qubit 
rotation, and can also be combined in such a way as to implement an 
entangling CNOT gate as well. This is due to the fact 
that the states which were used in Ref.~\cite{diVincenzo} for the 
CNOT are also present in the expansions of the logical states 
encoded into qudits having $d \geq$ 3.

In addition, the invariant $I_2^{(\alpha,\beta)}$ can also be used to perform 
the generalized exchange interaction between the states 
$\ket{p}^{(\alpha)}\ket{q}^{(\beta)}$ associated with particles 
$\alpha$ and $\beta$ since it has been shown in 
Ref.~\cite{Bishop/Byrd:09} that
\begin{equation}
\exp\left[-i(\pi/4)\sum_{j} \lambda_j \otimes \lambda_j \right]
\ket{\alpha \beta} = -i\exp(\pi i/2d)\ket{\beta \alpha},
\end{equation} 
for $\alpha,\beta = 1,2,\hdots,d.$

Clearly these are linear combinations of the two-body interactions
which are comprised of the invariants $I_2^{(\alpha,\beta)}$.  
Three-body and higher order interactions are less often experimentally
controllable, but are also, in principle, viable candidates for
quantum gates.  For example the logical $Y$ interaction for qudits is
proportional to $I_3$.


\section{Conclusions}
\label{sec:concl}

For quantum systems containing many particles, each having a number of
internal states, the system could be
in a vast array of possible states corresponding to a large Hilbert
space dimension.  The evolution of such states can be fairly simple
however, as in the case of a system undergoing collective
motion.  Such motions occur, for 
example, when $\lambda \gg R$ so that each particle feels the same
field.  If states, or subsystems, of a
collection of particles are invariant under collective motions, they
are decoherence-free, or noiseless with respect to any collective
operation, unitary or not.  This leads to the promising method for
error prevention--encoding in one of these subspaces to avoid
collective errors.  To take advantage of such an encoding for the
purposes of quantum information processing, one requires a complete
set of logical operations to be performed on these subsystems which is
compatible with the encoding.  We have provided a way in which to find
the set of Hamiltonians for this purpose.  

However, we also note that since collective motions commute with
the invariant operators we have presented here, the invariants may be
measured while the system undergoes these collective
errors.  This allows one to describe the system by the values of these
operators.  Indeed one of the original motivations for studying these
invariants was to find a complete set of commuting observables to
completely specify a quantum system.  (See for example
Ref.~\cite{Bohm:qmbook} and references therein.)  Not all of the
invariants presented here will commute with each other, but they each
commute with the collective motions.  A subset of these invariant
operators which also mutually commute will help provide a complete
set of commuting operators along with the energy and total angular
momentum.

Our work is quite general and can be applied to any set of $d$-state
systems undergoing collective motions.  Therefore, we have extended
the Dicke-state description explicitly to the general case leading the
way to the description of sets of particles undergoing collective
motions and their manipulation when the particles have more than two
internal states.


\begin{acknowledgments}
This material is based upon work supported by the National Science 
Foundation under Grant No. 0545798.
\end{acknowledgments}


\appendix


\section{The algebra of $SU(d)$}

\label{app:sunalg}

We have chosen the following convention for the normalization of 
the algebra of Hermitian matrices which are generators of $SU(d)$.  
\begin{equation}
\tr(\lambda_i\lambda_j) = 2\delta_{ij}.  
\end{equation}

The commutation and anticommutation relations of the matrices 
representing the basis for the Lie algebra can be summarized 
using the following equation:
\begin{equation}
\label{eq:larels}
\lambda_i \lambda_j = \frac{2}{d}\delta_{ij} + if _{ijk} \lambda_k 
                      + d_{ijk}\lambda_k,
\end{equation}
where here, and throughout this appendix, a sum over repeated 
indices is understood.  The sums are written explicitly 
for clarity only in a few cases.  

As with any Lie algebra we have the Jacobi identity:
\begin{equation}
\label{eq:jid}
f_{ilm}f_{jkl} + f_{jlm}f_{kil} + f_{klm}f_{ijl} =0.
\end{equation}
There is also a Jacobi-like identity,
\begin{equation}
\label{eq:jlid}
f_{ilm}d_{jkl} + f_{jlm}d_{kil} + f_{klm}d_{ijl} =0,
\end{equation}
which was given by Macfarlane, et al. \cite{Macfarlane}. 

The following identities, also provided in \cite{Macfarlane}, are
useful 
\begin{eqnarray}
d_{iik} &=& 0, \label{eq:Mac(2.7)}\\
d_{ijk}f_{ljk} &=& 0,  \label{eq:Mac(2.14)}\\
f_{ijk}f_{ljk} &=& d\delta_{il},  \label{eq:Mac(2.12)}\\
d_{ijk}d_{ljk} &=& \frac{d^2 - 4}{d}\delta_{il},  \label{eq:Mac(2.13)}
\end{eqnarray}
and
\begin{equation}
f_{ijm}f_{klm} = \frac{2}{d}(\delta_{ik}\delta_{jl} - \delta_{il}\delta_{jk}) 
                  + (d_{ikm}d_{jlm} - d_{jkm}d_{ilm}) \label{eq:Mac(2.10)}
\end{equation}
 and finally
\begin{eqnarray}
d_{piq}d_{qjr}f_{rkp} &=& \frac{d^2 - 4}{2d}f_{ijk}, \label{eq:Mac(2.17)}\\
d_{piq}d_{qjr}d_{rkp} &=& \frac{d^2 - 12}{2d}d_{ijk} \label{eq:Mac(2.18)}.
\end{eqnarray}
The proofs of these are fairly straight-forward, but we omit them here.



\begin{thebibliography}{23}
\expandafter\ifx\csname natexlab\endcsname\relax\def\natexlab#1{#1}\fi
\expandafter\ifx\csname bibnamefont\endcsname\relax
  \def\bibnamefont#1{#1}\fi
\expandafter\ifx\csname bibfnamefont\endcsname\relax
  \def\bibfnamefont#1{#1}\fi
\expandafter\ifx\csname citenamefont\endcsname\relax
  \def\citenamefont#1{#1}\fi
\expandafter\ifx\csname url\endcsname\relax
  \def\url#1{\texttt{#1}}\fi
\expandafter\ifx\csname urlprefix\endcsname\relax\def\urlprefix{URL }\fi
\providecommand{\bibinfo}[2]{#2}
\providecommand{\eprint}[2][]{\url{#2}}

\bibitem[{\citenamefont{{P. Zanardi and M. Rasetti}}(1997)}]{Zanardi:97c}
\bibinfo{author}{\bibnamefont{{P. Zanardi and M. Rasetti}}},
  \bibinfo{journal}{Phys. Rev. Lett.} \textbf{\bibinfo{volume}{79}},
  \bibinfo{pages}{{3306}} (\bibinfo{year}{1997}).

\bibitem[{\citenamefont{{L.-M Duan and G.-C. Guo}}(1998)}]{Duan:98}
\bibinfo{author}{\bibnamefont{{L.-M Duan and G.-C. Guo}}},
  \bibinfo{journal}{Phys. Rev. A} \textbf{\bibinfo{volume}{57}},
  \bibinfo{pages}{{737}} (\bibinfo{year}{1998}).

\bibitem[{\citenamefont{{D.A. Lidar, I.L. Chuang and K.B.
  Whaley}}(1998)}]{Lidar:PRL98}
\bibinfo{author}{\bibnamefont{{D.A. Lidar, I.L. Chuang and K.B. Whaley}}},
  \bibinfo{journal}{Phys. Rev. Lett.} \textbf{\bibinfo{volume}{81}},
  \bibinfo{pages}{{2594}} (\bibinfo{year}{1998}).

\bibitem[{\citenamefont{{E. Knill, R. Laflamme and L.
  Viola}}(2000)}]{Knill:99a}
\bibinfo{author}{\bibnamefont{{E. Knill, R. Laflamme and L. Viola}}},
  \bibinfo{journal}{Phys. Rev. Lett.} \textbf{\bibinfo{volume}{84}},
  \bibinfo{pages}{2525} (\bibinfo{year}{2000}).

\bibitem[{\citenamefont{{J. Kempe, D. Bacon, D.A. Lidar, and K.B.
  Whaley}}(2001)}]{Kempe:00}
\bibinfo{author}{\bibnamefont{{J. Kempe, D. Bacon, D.A. Lidar, and K.B.
  Whaley}}}, \bibinfo{journal}{Phys. Rev. A} \textbf{\bibinfo{volume}{63}},
  \bibinfo{pages}{042307} (\bibinfo{year}{2001}).

\bibitem[{\citenamefont{{D.A. Lidar, D. Bacon, J. Kempe, and K.B.
  Whaley}}(2001)}]{Lidar:00a}
\bibinfo{author}{\bibnamefont{{D.A. Lidar, D. Bacon, J. Kempe, and K.B.
  Whaley}}}, \bibinfo{journal}{Phys. Rev. A} \textbf{\bibinfo{volume}{63}},
  \bibinfo{pages}{022306} (\bibinfo{year}{2001}).

\bibitem[{\citenamefont{{D.A. Lidar and K.B. Whaley}}(2003)}]{Lidar/Whaley:03}
\bibinfo{author}{\bibnamefont{{D.A. Lidar and K.B. Whaley}}}, in
  \emph{\bibinfo{booktitle}{{Irreversible Quantum Dynamics}}}
  (\bibinfo{publisher}{{Springer-Verlag}}, \bibinfo{address}{{Berlin}},
  \bibinfo{year}{2003}).

\bibitem[{\citenamefont{{M. S. Byrd, L.-A. Wu and D. A.
  Lidar}}(2004)}]{Byrd/etal:pqe04}
\bibinfo{author}{\bibnamefont{{M. S. Byrd, L.-A. Wu and D. A. Lidar}}},
  \bibinfo{journal}{J. Mod. Optics} \textbf{\bibinfo{volume}{51}},
  \bibinfo{pages}{{2449}} (\bibinfo{year}{2004}).

\bibitem[{\citenamefont{Dicke}(1954)}]{Dicke:54}
\bibinfo{author}{\bibfnamefont{R.}~\bibnamefont{Dicke}},
  \bibinfo{journal}{Phys. Rev.} \textbf{\bibinfo{volume}{93}},
  \bibinfo{pages}{99} (\bibinfo{year}{1954}).

\bibitem[{par()}]{partnote}
\bibinfo{note}{{We will refer to the constituents as particles although we
  emphasize that the constituents need not be individual particles. They could
  be sets of particles, collections of subsystems, etc.}}

\bibitem[{\citenamefont{Prevedel et~al.}(2009)\citenamefont{Prevedel,
  Cronenberg, Tame, Paternostro, Walther, Kim, and
  Zeilinger}}]{Prevedel/etal:09}
\bibinfo{author}{\bibfnamefont{R.}~\bibnamefont{Prevedel}},
  \bibinfo{author}{\bibfnamefont{G.}~\bibnamefont{Cronenberg}},
  \bibinfo{author}{\bibfnamefont{M.~S.} \bibnamefont{Tame}},
  \bibinfo{author}{\bibfnamefont{M.}~\bibnamefont{Paternostro}},
  \bibinfo{author}{\bibfnamefont{P.}~\bibnamefont{Walther}},
  \bibinfo{author}{\bibfnamefont{M.~S.} \bibnamefont{Kim}}, \bibnamefont{and}
  \bibinfo{author}{\bibfnamefont{A.}~\bibnamefont{Zeilinger}},
  \bibinfo{journal}{Phys. Rev. Lett.} \textbf{\bibinfo{volume}{103}},
  \bibinfo{pages}{020503} (\bibinfo{year}{2009}).

\bibitem[{\citenamefont{Wieczorek et~al.}(2009)\citenamefont{Wieczorek,
  Krischek, Kiesel, Michelberger, T\'{o}th, and
  Weinfurter}}]{Wieczorek/etal:09}
\bibinfo{author}{\bibfnamefont{W.}~\bibnamefont{Wieczorek}},
  \bibinfo{author}{\bibfnamefont{R.}~\bibnamefont{Krischek}},
  \bibinfo{author}{\bibfnamefont{N.}~\bibnamefont{Kiesel}},
  \bibinfo{author}{\bibfnamefont{P.}~\bibnamefont{Michelberger}},
  \bibinfo{author}{\bibfnamefont{G.}~\bibnamefont{T\'{o}th}}, \bibnamefont{and}
  \bibinfo{author}{\bibfnamefont{H.}~\bibnamefont{Weinfurter}},
  \bibinfo{journal}{Phys. Rev. Lett.} \textbf{\bibinfo{volume}{103}},
  \bibinfo{pages}{020504} (\bibinfo{year}{2009}).

\bibitem[{\citenamefont{H\"{a}rk\"{o}nen
  et~al.}(2009)\citenamefont{H\"{a}rk\"{o}nen, Plastina, and
  Maniscalco}}]{Harkonen/etal:09}
\bibinfo{author}{\bibfnamefont{K.}~\bibnamefont{H\"{a}rk\"{o}nen}},
  \bibinfo{author}{\bibfnamefont{F.}~\bibnamefont{Plastina}}, \bibnamefont{and}
  \bibinfo{author}{\bibfnamefont{S.}~\bibnamefont{Maniscalco}},
  \bibinfo{journal}{Phys. Rev. A} \textbf{\bibinfo{volume}{80}},
  \bibinfo{pages}{033841} (\bibinfo{year}{2009}).

\bibitem[{\citenamefont{M.~Kiffner and Keitel}(2007)}]{Kiffner/etal:07}
\bibinfo{author}{\bibfnamefont{J.~E.} \bibnamefont{M.~Kiffner}}
  \bibnamefont{and} \bibinfo{author}{\bibfnamefont{C.}~\bibnamefont{Keitel}},
  \bibinfo{journal}{Phys. Rev. A} \textbf{\bibinfo{volume}{75}},
  \bibinfo{pages}{032313} (\bibinfo{year}{2007}).

\bibitem[{\citenamefont{{C.A. Bishop and M.S. Byrd}}(2008)}]{Bishop/Byrd:08}
\bibinfo{author}{\bibnamefont{{C.A. Bishop and M.S. Byrd}}},
  \bibinfo{journal}{Phys. Rev. A} \textbf{\bibinfo{volume}{77}},
  \bibinfo{pages}{{012314}} (\bibinfo{year}{2008}).

\bibitem[{\citenamefont{D.B.~Hume and Wineland}(2009)}]{Hume/etal:09}
\bibinfo{author}{\bibfnamefont{T.}~\bibnamefont{D.B.~Hume},
  \bibfnamefont{C.W.~Chou}} \bibnamefont{and}
  \bibinfo{author}{\bibfnamefont{D.}~\bibnamefont{Wineland}},
  \bibinfo{journal}{Phys. Rev. A} \textbf{\bibinfo{volume}{80}},
  \bibinfo{pages}{052302} (\bibinfo{year}{2009}).

\bibitem[{\citenamefont{{B. Gruber and L. O'Raifeartaigh}}(1964)}]{Gruber:64}
\bibinfo{author}{\bibnamefont{{B. Gruber and L. O'Raifeartaigh}}},
  \bibinfo{journal}{J. Math. Phys.} \textbf{\bibinfo{volume}{5}},
  \bibinfo{pages}{{1796}} (\bibinfo{year}{1964}).

\bibitem[{\citenamefont{{J\"urgen Fuchs and Christoph
  Schweigert}}(1997)}]{Fuchs/Schweigert}
\bibinfo{author}{\bibnamefont{{J\"urgen Fuchs and Christoph Schweigert}}},
  \emph{\bibinfo{title}{{Symmetries, Lie Algebras and Representations}}}
  (\bibinfo{publisher}{{Cambridge University Press}}, \bibinfo{year}{1997}).

\bibitem[{\citenamefont{{M.S. Byrd}}(2006)}]{Byrd:06}
\bibinfo{author}{\bibnamefont{{M.S. Byrd}}}, \bibinfo{journal}{Phys. Rev. A}
  \textbf{\bibinfo{volume}{73}}, \bibinfo{pages}{{032330}}
  (\bibinfo{year}{2006}).

\bibitem[{\citenamefont{{C.A. Bishop and M.S. Byrd}}(2009)}]{Bishop/Byrd:09}
\bibinfo{author}{\bibnamefont{{C.A. Bishop and M.S. Byrd}}},
  \bibinfo{journal}{{J. Phys. A: Math. Theor.}} \textbf{\bibinfo{volume}{42}},
  \bibinfo{pages}{{055301}} (\bibinfo{year}{2009}).

\bibitem[{\citenamefont{{D.P. DiVincenzo}}(1995)}]{diVincenzo}
\bibinfo{author}{\bibnamefont{{D.P. DiVincenzo}}}, \bibinfo{journal}{Science}
  \textbf{\bibinfo{volume}{270}}, \bibinfo{pages}{255} (\bibinfo{year}{1995}).

\bibitem[{\citenamefont{{A. Bohm}}(1993)}]{Bohm:qmbook}
\bibinfo{author}{\bibnamefont{{A. Bohm}}}, \emph{\bibinfo{title}{{Quantum
  Mechanics: Foundations and Applications, 3rd Ed., Chapter 5}}}
  (\bibinfo{publisher}{{Springer-Verlag}}, \bibinfo{address}{{New York, New
  York}}, \bibinfo{year}{1993}).

\bibitem[{\citenamefont{{A.J. Macfarlane, A. Sudbery and P.H.
  Weisz}}(1968)}]{Macfarlane}
\bibinfo{author}{\bibnamefont{{A.J. Macfarlane, A. Sudbery and P.H. Weisz}}},
  \bibinfo{journal}{{Commun. Math. Phys.}} \textbf{\bibinfo{volume}{11}},
  \bibinfo{pages}{{77}} (\bibinfo{year}{1968}).

\end{thebibliography}


\end{document}